\documentclass[aps,pre,superscriptaddress,showpacs,floats,twocolumn,amssymb,floatfix]{revtex4-1}

\input epsf
\usepackage{bm}
\usepackage{amsmath,amsfonts}
\usepackage{graphicx}
\usepackage{color}
\usepackage[usenames,dvipsnames,svgnames,table]{xcolor}
\usepackage{hyperref}
\usepackage{setspace}
\usepackage{soul}
\usepackage{accents}
\usepackage{mathtools}

\newcommand{\beq}{\begin{equation}}
\newcommand{\eeq}{\end{equation}}
\newcommand{\beqa}{\begin{eqnarray}}
\newcommand{\eeqa}{\end{eqnarray}}
\newcommand{\beqan}{\begin{eqnarray*}}
\newcommand{\eenan}{\end{eqnarray*}}

\newcommand{\Dop}{{D_{{\hat{p}}}}}
\newcommand{\Dov}{{D_{{\hat{v}}}}}
\newcommand{\Dow}{{D_{{\hat{\omega}}}}}
\newcommand{\CU}{{ U}}

\newcommand{\CR}{{\cal R}}

\newcommand{\ratepv}{\gamma}
\newcommand{\ratew}{\gamma_\omega}

\newcommand{\jz}{{\cal J}_z}

\newcommand{\px}{{\cal P}_x}
\newcommand{\py}{{\cal P}_y}
\newcommand{\pz}{{\cal P}_z}

\newcommand{\ppx}{{\cal P}_x^\perp}
\newcommand{\ppy}{{\cal P}_y^\perp}
\newcommand{\ppz}{{\cal P}_z^\perp}

\newcommand{\km}{ {\cal K}}

\newcommand{\freq}{{f}}
\newcommand{\ws}{\omega_z}

\newcommand\defeq{:{\kern -0.5em}=}

\newcommand{\unitp}{\hat{p}}
\newcommand{\unitv}{\hat{v}}
\newcommand{\unitw}{\hat{\omega}}
\newcommand{\unitxi}{\hat{\xi}}

\newcommand{\unitx}{\hat{x}}
\newcommand{\unity}{\hat{y}}
\newcommand{\unitz}{\hat{z}}

\newcommand{\Deff}{ D_{\mbox{\scriptsize eff}} }

\begin{document}

\addtolength{\topmargin}{10pt}

\title{
Kinematic matrix theory and universalities in self-propellers and active swimmers}
\author{Amir Nourhani} 
\email{nourhani@psu.edu}
\affiliation{Department of Physics, The Pennsylvania State University, University Park, PA 16802}
\author{Paul E. Lammert}
\affiliation{Department of Physics, The Pennsylvania State University, University Park, PA 16802} 
\author{Ali Borhan}
\affiliation{Department of Chemical Engineering, The Pennsylvania State University, University Park, PA 16802} 
\author{Vincent H. Crespi}
\affiliation{Department of Physics, The Pennsylvania State University, University Park, PA 16802}
\affiliation{Department of Chemistry, The Pennsylvania State University, University Park, PA 16802}
\affiliation{Department of Materials Science and Engineering, The Pennsylvania State University, University Park, PA 16802}

\begin{abstract}
We describe an efficient and parsimonious matrix-based theory for studying the ensemble behavior of self-propellers and active swimmers, such as nanomotors or motile bacteria, that are typically studied by differential-equation-based Langevin or Fokker-Planck formalisms. The kinematic effects for elementary processes of motion are incorporated into a matrix, called the ``kinematrix'', from which we immediately obtain correlators and the mean and variance of angular and position variables (and thus effective diffusivity) by simple matrix algebra. The kinematrix formalism enables us recast the behaviors of a diverse range of self-propellers into a unified form, revealing universalities in their ensemble behavior in terms of new emergent time scales. Active fluctuations and hydrodynamic interactions can be expressed as an additive composition of separate self-propellers. 
\end{abstract}
\maketitle 

\section{INTRODUCTION}
\begin{figure}[b]
\begin{center}
\includegraphics[width=3.1 in]{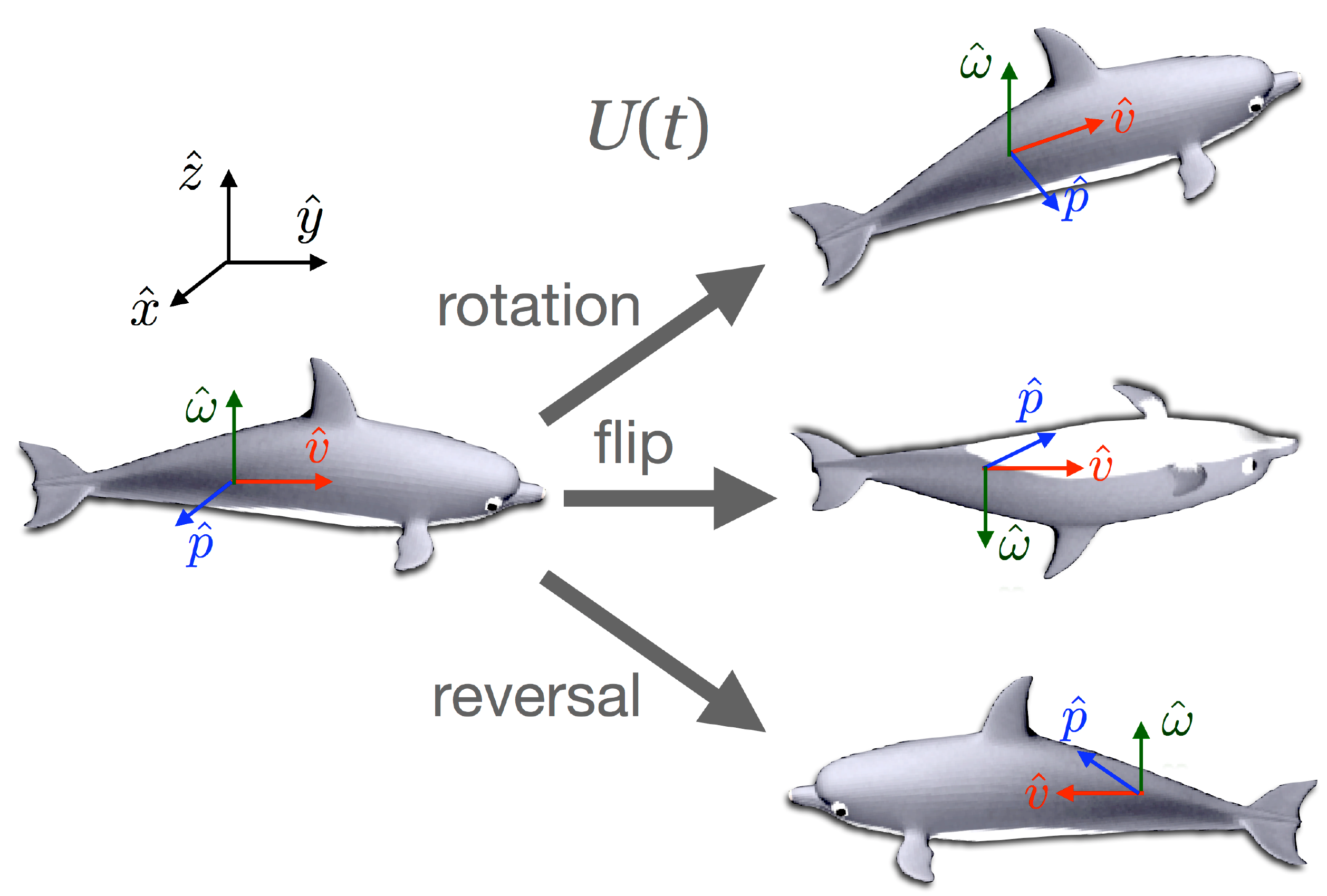}
\end{center}
\vspace{-15pt}
\caption{
At $t=0$ the laboratory frame and body frame coincide. During $[0,dt)$ a self-propeller can experience small turns about $\unitw$ due to  deterministic rotation, stochastic orientational diffusion, or tumbling; it can also flip about $\unitv$ or reverse the direction of motion. The operator $U(t)$ represents this evolution.
\label{fig:dolphin}}
\end{figure}

Self-propellers are a wide class of far-from-equilibrium systems including motile cells and bacteria 
\cite{
Berg:2000p601,    
Zaburdaev2011PRL208103,                  
Bodeker:2010p593, 
Li:2008p594,
Li:2011p600,
Bodeker2010EPL28005,
Peruani:2007p602,
Campos2010JTB526,
Li-ProcNatlAcad:2008p614,     
PNAS-1995-Frymier-6195-9,  
Lauga:2006p606,                     
Kudo-FEMS-2005p221,           
Riedel-Science:p300,               
FLM:378272,                            
PhysRevE.73.021505,            
Erglis:2007p596,                      
Cebers:2011p639,
Selmeczi:2005p599,
Selmeczi:2008p603},                    
artificial nanomotors 
\cite{
Howse2007PRL048102,
Paxton:2004p183, 
Paxton:2005p154, 
Paxton:2006p37, 
Wang:2009p109, 
Wang:2010p106, 
Gibbs:2011p546, 
Ozin:2005p173, 
Mirkovic:2010p642, 
Mirkovic:2010p126, 
Ebbens:2010p86,
Golestanian2009PRL188305},   
aquatic swimmers \cite{
Gautrais2009JMB429,
citeulike:11429599,
Ordemann2003260,
Mach2007BMB539},                        
insects, birds and other animals 
\cite{Edwards:2007p598, 
COUZIN20021, 
Niwa1994123, 
Bazazi:2011p649, 
Bazazi2008735, 
Strefler:2009p605}, 
and even pedestrians 
\cite{RevModPhys.73.1067}. 
Their motion naturally decomposes into distinct elementary processes such as deterministic  translation and rotation, plus stochastic components such as orientational diffusion, flipping about an axis, or tumbling (Fig.~\ref{fig:dolphin}). The phenomenological kinematic parameters of these elementary processes are typically obtained by comparing
experimental observations to theoretical models that are developed using Langevin or Fokker-Planck differential equations~\cite{Romanczuk:2011p87,
Romanczuk:2012p624, 
Dunkel:2009p646,
Takagi:2013p645, 
VanTeeffelen:2008p643,
Weber:2011p644,
Nourhani2013p050301,
Bodeker:2010p593,
Li:2008p594,
Li:2011p600,
Bodeker2010EPL28005,
Peruani:2007p602,
Campos2010JTB526,
Strefler:2009p605,
Ebbens:2010p589,
RevModPhys.73.1067,
Lobaskin2008EPJST157,
Schweitzer1998PRL5044,
Si2012PA3054,
Cates2013EPL20010,
Romanczuk2009PRL010602,
Chepizhko2013PRL160604}.
These mathematical formalisms 
grow more cumbersome as the number of elementary processes increases, with distinct processes often being treated in a non-uniform manner by a menu of methods. Here we introduce an alternative {\it kinematrix theory} based on matrix algebra, which treats all elementary motive processes on an equal footing and remains tractable for complex motor behavior. 
By inspection, we compile the kinematic effects of the elementary processes into a matrix, called the {\em kinematrix}, from which we immediately obtain the ensemble behavior of self-propellers by simple matrix algebra. The kinematrix consolidates the behavior of many classes of self-propellers~\cite{FLM:378272, PhysRevE.73.021505, Erglis:2007p596, Cebers:2011p639, VanTeeffelen:2008p643, Nourhani2013p050301, Weber:2011p644, Friedrich:2008p610, Ebbens:2010p589, Polin-Science:p487, Bennett:2013p586} into a single unified form with newly emergent composite timescales, thus revealing universalities in the ensemble behavior of diverse self-propellers that had previously been considered `different' systems. The analytical and computational simplicity of the kinematrix should also facilitate further advances in the analysis of large complex datasets, such as the inverse problem of extracting the correct elementary motive processes from complex trajectory data. The effects of active fluctuations and rotation-translation coupling at the level of ensemble properties can be recast as effectively additive contributions arising from independent self-propellers. 

The key to the kinematrix approach is a body-frame description of the motion. A self-propeller, while translating, may rotate deterministically due to structural imperfection
\cite{Takagi:2013p645,
Laocharoensuk:2008p64, 
Dhar:2006p117}, 
hydrodynamic interaction with a substrate 
\cite{
Li-ProcNatlAcad:2008p614, 
PNAS-1995-Frymier-6195-9, 
Lauga:2006p606,
Kudo-FEMS-2005p221, 
Riedel-Science:p300},
or purposeful engineering 
\cite{
nourhani2013p062317,
Wang:2009p80,
FournierBidoz:2005p95,
Qin:2007p165,
Gibbs:2011p96,
Gibbs:2010p68,
Gibbs:2009p85,
Valadares:2010p139,                  
FournierBidoz:2005p95};            
they may also suffer stochastic events that influence the direction of motion such as 
tumbling or orientational diffusion. We build an empirical body frame that is anchored in dynamical (rather than geometrical) properties: the self-propeller translates at velocity ${\bm v}= v \unitv$ 
while rotating at angular velocity ${\bm \omega} = \omega \unitw\,$ at a position ${\bm p} = {v \over \omega} \unitp$ with respect to the center of its instantaneous circular orbit, thus yielding an orthonormal triple $[\unitp, \unitv, \unitw]$ as a right-handed empirical body frame fixed to the self-propeller. We choose a fixed laboratory frame $[\unitx, \unity, \unitz]$ that coincides with $[\unitp, \unitv, \unitw]$ at $t=0$. The orientation of body frame at time $t$ is related to its initial orientation by a propagator $U(t)$ such that $\unitxi(t) = \CU(t) \, \unitxi(0)$ for $\unitxi$ being any of $\unitp$, $\unitv$ or $\unitw$. The propagator represents the net deterministic and stochastic rotation of the body frame from $0 \rightarrow t$; from its ensemble average $\langle U(t)\rangle$ we can obtain pair correlators, finite-time average displacement, finite-time mean-square displacement, and asymptotic effective diffusivity. The stochastic component of the dynamics leads to an ensemble of possible body frame orientations at time $t$, and each orientation connects to the initial orientation though an ensemble of different paths. The core of kinematrix theory is to overcome the difficulty of obtaining $\langle \CU (t) \rangle$ over these ensembles by turning $\langle \CU (t) \rangle$ into the product of ensemble averages of independent incremental rotations in the body frame, as explained below.

\section{THEORY FORMULATION}
Divide the timeline $[0,t)$~into infinitesimal increments $dt = t/n$ (large integer $n$) with endpoints $t_i \equiv i \, dt$ (integer $i$).  
With $\CU_i \equiv \CU(t_i)$, and $\CR_i$ as the net  rotation in the laboratory frame during the infinitesimal interval $[t_i,t_{i+1})$,  the propagator takes the  recursive form
$\CU_n = \CR_{n-1} \CU_{n-1}$. Now, we transform the rotations $\CR_i$ in the laboratory frame into  rotations in the body frame
$\tilde{\CR}_i = \CU_{i}^{-1} \CR_i \CU_{i}$,
yielding,
$\CU_n = \CR_{n-1} \CU_{n-1} = \CU_{n-1} \tilde{\CR}_{n-1} = \tilde{\CR}_0  \tilde{\CR}_1 \cdots  \tilde{\CR}_{n-1}$.
For processes with negligible correlation time, the $\tilde{\CR}_i$'s are independent
and identically distributed, $\langle\tilde{\CR}_i\rangle = \langle\tilde{\CR}_0\rangle$,
resulting in
\beq
\langle \CU(t) \rangle  
\simeq
\langle \tilde{\CR}_0 \rangle^{t/dt}. 
\label{eq:exppropag}
\eeq
The net rotation $\tilde{\CR}_0 = \tilde{\cal R}_0^{(1)} \cdots \tilde{\cal R}_0^{(N)}$ is the product of $N$ independent elementary processes $\tilde{\CR}_0^{(j)}$ (such as flipping, tumbling or orientational diffusion) involved in self-propeller's motion during [0,dt). Expanding their expectations to first order in $dt$, $\langle \tilde{\CR}_0^{(j)} \rangle = {\cal I} - \km^{(j)} \, dt + {\cal O}(dt^2)$ for $j=1,\cdots,N$, we obtain 
\beq
\langle \tilde{\CR}_0 \rangle = {\cal I} - \km \, dt + {\cal O}(dt^2).
\label{eq:dynmat}
\eeq
We call $\km = \km^{(1)} + \cdots + \km^{(N)}$ the {\it kinematrix}. It is the sum of the first-order contributions $\km^{(i)}$ of elementary rotations and contains the kinematic effects of all such processes. In the limit $t \gg dt$, Eqs. (\ref{eq:exppropag}) and (\ref{eq:dynmat}) yield
\beq
\langle \CU(t) \rangle = e^{-\km  t}.
\label{eq:propagator}
\eeq
Using the initial condition $[\unitx, \unity, \unitz] \equiv [\unitp(0), \unitv(0), \unitw(0)]$ and $\unitxi(t) = \CU(t) \, \unitxi(0)$ for $\unitxi$ any of $\unitp, \unitv$, or $\unitw$, 
we obtain the correlators for linear ($v \unitv$) and angular ($\omega\unitw$) velocities 
\begin{align}
&
C_{{\bm v}{\bm v}}(t) = 
\langle {\bm v}(0)  \cdot   {\bm v}(t) \rangle
 =
\langle v(0)    v(t) \rangle
 \left[ e^{-\km t}\right]_{22}
  \label{eq:velcorrel}
\\
&
C_{{\bm \omega}{\bm \omega}}(t) = 
\langle {\bm \omega}(0)  \cdot   {\bm \omega}(t) \rangle
=	
\langle \omega(0)  \omega(t) \rangle
 \left[ e^{-\km t}\right]_{33}.
  \label{eq:angularvelcorrel}
\end{align}
The subscripts on $\left[ e^{-\km t}\right]$ identify matrix elements. The off-diagonal elements of $e^{-\km t}$ give the correlators between different directions $\unitp$, $\unitv$, $\unitw$. 

The magnitude and direction of the velocity may fluctuate as a result of random  disturbances, but since the sources of these fluctuations generally differ, the fluctuations of $v$ and $\unitv$ are typically independent. Before discussing the effect of active speed fluctuations, we explain the formalism with a speed $v(t)$ fluctuating weakly around a mean $\langle v(t) \rangle = \bar{v}$ with negligible correlation time, so that $\langle v(0) v(t) \rangle \simeq  \bar{v}^2$ (the common assumption of constant speed \cite{Ebbens:2010p589, Friedrich:2008p610, Weber:2011p644, VanTeeffelen:2008p643, Takagi:2013p645, Nourhani2013p050301, B911365G} is a special case). Thus, the ensemble average and mean squared displacements are
\begin{align}
&\langle \Delta  {\bm r}(t) \rangle  
\! =\bar{v}\! \begingroup\textstyle\int\endgroup_0^t  e^{-\km  t} \! \cdot\!  \unitv(0)\, dt^\prime = \bar{v} \,\km^{-1}\! \left({\cal I}\! -\! e^{-\km t} \right)\! \cdot \!\unitv(0)
\label{eq:displacementvec}
\\[.5 em]
& \langle|\Delta{\bm r}(t)|^2\rangle 
 =  2 \bar{v}^2 \left[ t\, \km^{-1}  - \km^{-1}\km^{-1} \left({\cal I} - e^{-\km t} \right) \right]_{22}.
\label{eq:MSD1}
\end{align}
The interplay of deterministic and stochastic dynamics creates effective diffusion at long times; the observed diffusivity of a self-propeller is the sum of its effective and passive Brownian diffusivities. The effective diffusivity $\Deff  = {1\over 2d} \lim_{t\to\infty} t^{-1} \langle|\Delta{\bm r}(t)|^2\rangle$ in $d$ dimensional space is
\begin{align}
\Deff = {\bar{v}^2 \over d}\left[\km^{-1}\right]_{22} = {\bar{v}^2 \over d} \, {\km_{11} \km_{33} - \km_{13}\km_{31} \over \mbox{det} \km}.
\label{eq:DeffMinverse}
\end{align}
If motion is strictly restricted to a plane such that $\mbox{det} \, \km=0$,
we make the replacement $\km \to \km + \varepsilon {\cal I}$, calculate Eqs.  (\ref{eq:velcorrel})--(\ref{eq:DeffMinverse}), and take $\varepsilon\to0$ at the end.

Unsteady operation of the self-propeller's engine leads to speed fluctuations $\delta v(t)$. Since $\langle v(t) v(0)\rangle = \overline{v}^2 + \langle \delta v(t)\, \delta v(0)\rangle$, as far as velocity correlations (\ref{eq:velcorrel}), mean-square displacement (\ref{eq:MSD1}), and effective diffusivity (\ref{eq:DeffMinverse}) are concerned, the velocity fluctuations $\delta v(t)\, \hat{v}(t)$ make an uncorrelated additive contribution. For models~\cite{Peruani:2007p602} where $\langle \delta v(t)\, \delta v(0)\rangle = [\langle v^2\rangle - \overline{v}^2]e^{-\kappa_v t}$ such contributions are equal to the corresponding properties of a self-propeller with mean speed $\sqrt{\langle v^2\rangle - \overline{v}^2}$ and kinematrix $\km^\prime = \km + \kappa_v{\cal I}$.

\section{ELEMENTARY PROCESSES}
Equations (\ref{eq:velcorrel})--(\ref{eq:DeffMinverse}) show how important physical quantities follow immediately from the kinematrix  $\km$.
Next, we show how to build $\km$ from the contributions $\km^{(j)}$ of elementary processes by inspection of motion during the interval $[0,dt)$.  We write $\km$ in terms of a linear combination of matrices $\mathcal{J}_k$, $\mathcal{P}_k$ and $\mathcal{P}^\perp_k$.
$\mathcal{J}_k$ is the generator of infinitesimal rotation about an axis $\hat{k}$, 
\beq
\mathcal{J}_x
\!
=
\!\!
\left[
\!
 \begin{array}{ccc} 
  0 &  0 & 0  \\
  0 &  0 & -1  \\ 
  0 &  1  &  0 
 \end{array} 
\! \!
\right]
 \!\!
, 
\ 
\mathcal{J}_y
\!
=
\!\!
\left[
\!\!
 \begin{array}{ccc} 
  0 &  0 & 1  \\
  0 &  0 & 0  \\ 
  -1 &  0  &  0 
 \end{array} 
 \!
 \right]  
 \!\!
 ,
 \ 
\mathcal{J}_z
\!
=
\!\!
\left[
\!
 \begin{array}{ccc} 
  0 &  -1 & 0  \\
  1  &  0 & 0  \\ 
  0 &  0  &  0 
 \end{array} 
 \!
 \right]
 \!\!
,
 \label{eq:Jmatrices}
\eeq
${\cal P}_{k}$  is the orthogonal projection onto the $k$-th coordinate,  
\beq
\mathcal{P}_x
\!
=
\!\!
\left[
 \begin{array}{ccc} 
  1 &  0 & 0  \\
  0 &  0 & 0  \\ 
  0 &  0  &  0 
 \end{array} 
 \right] 
 \!\!
, 
\
\mathcal{P}_y
\!
=
\!\!
\left[
 \begin{array}{ccc} 
  0 &  0 & 0  \\
  0 &  1 & 0  \\ 
  0 &  0  &  0 
 \end{array} 
 \right]  
 \!\!
 , 
 \
\mathcal{P}_z
\!
=
\!\!
\left[
 \begin{array}{ccc} 
  0 &  0 & 0  \\
  0  &  0 & 0  \\ 
  0 &  0  &  1 
 \end{array} 
 \right]
 \!\!,   
 \label{eq:Pmatrices}
\eeq
and ${\cal P}_{k}^\perp = {\cal I} -{\cal P}_{k}$ is the projection onto the plane perpendicular to $\hat{k}$,
\beq
\mathcal{P}_x^\perp
\!
=
\!\!
\left[
 \begin{array}{ccc} 
  0 &  0 & 0  \\
  0 &  1 & 0  \\ 
  0 &  0  &  1 
 \end{array} 
 \!
 \right]
 \!\!
, 
\
\mathcal{P}_y^\perp
\!
=
\!\!
\left[
 \begin{array}{ccc} 
  1 &  0 & 0  \\
  0 &  0 & 0  \\ 
  0 &  0  &  1 
 \end{array} 
 \!
 \right] 
 \!\!
 ,
 \
\mathcal{P}_z^\perp
\!
=
\!\!
\left[
 \begin{array}{ccc} 
  1 &  0 & 0  \\
  0  &  1 & 0  \\ 
  0 &  0  &  0 
 \end{array} 
 \!
 \right]  
 \!
  .
 \label{eq:PPmatrices}
\eeq 
We study four elementary processes: tumbles, flips, deterministic rotation, and orientational diffusion; these four processes cover a wide variety of self-propeller dynamics from artificial nanomotors to motile single-cell organisms and macroscopic movers. A tumble is a sudden rotation of angle $\theta$ with distribution $P(\theta)$ about an axis $k$, occurring at a rate $\freq$. 
We model tumbles by a Poisson process $\tilde{\cal R}_0^{\mbox{\scriptsize tumble}} = {\cal I} + s_0 \left[ (\cos \theta-1) {\cal P}^\perp_k + \sin\theta {\cal J}_k \right]$ where $s_0$ is $0$ or $1$ with probabilities $(1-\freq) dt$ or $\freq dt$, respectively. The contribution of tumbling is then
\beq 
\km^{\mbox{\scriptsize tumble}}_{\hat{k}} = \freq (1 - \langle \cos\theta\rangle_{\!_P} ) {\cal P}_k^\perp - \freq \langle \sin\theta \rangle_{\!_P} {\cal J}_k
\label{eq:Dtumble}
\eeq
where the averages $\langle \cos\theta\rangle_{\!_P}$ and $\langle \sin\theta\rangle_{\!_P}$ are with respect to $P(\theta)$. Flipping is tumbling by $\theta = \pi$ about an axis $k$:
\beq 
\km^{\mbox{\scriptsize flip}}_{\hat{k}} = 2 \freq {\cal P}_k^\perp.
\label{eq:Dflip}
\eeq
Deterministic rotation at angular speed $\omega$ about axis $k$ during $dt$ is described by $e^{\omega dt {\cal J}_k} = {\cal I} - (-\omega {\cal J}_k) dt + {\cal O}(dt^2)$, thus 
\beq 
\km^{\mbox{\scriptsize det}}_{\hat{k}} = - \omega {\cal J}_k.
\label{eq:Ddet}
\eeq
Orientational diffusion with diffusivity $D_{\hat{k}}$ about an axis $k$ during $dt$ is described by $\tilde{\cal R}_0^{\mbox{\scriptsize ort}} = e^{d\phi {\cal J}_k}$ where the incremental rotational angle $d\phi$ is a Wiener process of mean zero and standard deviation $\sqrt{2 D_{\hat{k}} dt}$. The contribution of 
$\langle \tilde{\cal R}_0^{\mbox{\scriptsize ort}} \rangle = {\cal I} -  D_{\hat{k}} {\cal P}_k^\perp dt + {\cal O}(dt^2)$ to the kinematrix is 
\beq
\km^{\mbox{\scriptsize ort}}_{\hat{k}} = D_{\hat{k}} {\cal P}_k^\perp.
\label{eq:Dort}
\eeq

\section{FOUR SELF-PROPELLER SCENARIOS}
Having established the formalism, we now demonstrate how the elementary processes of Eqs. (\ref{eq:Dtumble}--\ref{eq:Dort}) combine to describe four different classes of self-propellers, building $\km$ by inspection. We start by illustrating the application of our formalism to  the previously studied classes of diffusing-and-flipping self-propellers and magnetotactic bacteria in a rotating magnetic field, calculating the asymptotic $\Deff$ (\ref{eq:DeffMinverse}) 
as an example of
ensemble properties (\ref{eq:velcorrel})--(\ref{eq:DeffMinverse}). Next we study run-and-tumble motion via our continuous model -- rather than traditional discrete random 
walks~\cite{
Seuront2004Zoological498,
Selmeczi:2005p599,
Selmeczi:2008p603,
citeulike:11429599,
Wu2000EM115} 
-- and demonstrate how to incorporate speed fluctuations. Last, we analyze a  3D swimmer and discuss the effect of rotation-translation coupling. After discussing these individual classes of self-propellers, we recast the kinematrices into a form that covers all motor types; the 
clarity of the formalism reveals new emergent time scales with universal short and long time 
behavior. 
In building $\km$ for each class,  keep in mind that we inspect the motion during $[0,dt)$, hence, ${\cal P}^\perp_{p}=\ppx$,  ${\cal P}^\perp_{v}=\ppy$,  ${\cal P}^\perp_{\omega}=\ppz$, and ${\cal J}_{\omega} = \jz$ in Eqs. (\ref{eq:Dtumble})--(\ref{eq:Dort}).

\noindent {\bf Diffusion-Flip (DF) self-propellers} include biological ``circle swimmers'' as well as artificial nanomotors~\cite{Takagi:2013p645, VanTeeffelen:2008p643, Nourhani2013p050301, Weber:2011p644, Friedrich:2008p610, Ebbens:2010p589}. Steric hinderance from a two dimensional (2D) planar substrate resists free orientational diffusion about any axis parallel to the substrate and forces the self-propeller to perform only sudden flips about $\unitv$. While translating and rotating in a 2D plane, the swimmer undergoes orientational diffusion with diffusivity $\Dow$ about $\unitw$ perpendicular to the substrate while flipping with frequency $\freq$ about $\unitv$ and changing its rotation chirality.
The  contributions to $\km$ are $-\omega \jz$ for deterministic rotation about $\unitw$, $\Dow \ppz$ for orientational diffusion about  $\unitw$, and $2 \freq \ppy$ for flipping about $\unitv$, so that $\km_{\mbox{\scriptsize 2D}} =  -  \omega \jz + \Dow\! \ppz + 2 \freq \ppy$. Eq. (\ref{eq:DeffMinverse}) then yields
(see Sec. 1 of the Appendix)
\beq
D_{\mbox{\scriptsize eff,2D}} = {\bar{v}^2 \over 2}
{ \Dow + 2 \freq \over \omega^2 + \Dow  \left(\Dow + 2 \freq\right)}
\label{eq:Deffective2Dflip}
\eeq
For a fast flipping rotor ($\freq \gg \omega$), the rotation rapidly averages out and the rotor acts like a linear motor 
($D_{\mbox{\scriptsize eff,2D}}^{\mbox{\scriptsize linear}} = {\bar{v}^2 \over 2 \Dow}$).

\noindent {\bf Magnetotactic bacteria (MB)} can move near a substrate and rotate in synchrony with a rotating magnetic field at angular speed $\omega$~\cite{FLM:378272, PhysRevE.73.021505, Erglis:2007p596, Cebers:2011p639}. The rotation contributes $- \omega {\cal J}_z$. The trajectory is a set of U-shaped segments due to occasional reversals of $\unitv$ with frequency $\freq$ while preserving the chirality of the orbit (contributing $2\freq \ppz$), plus orientational diffusion about $\unitw$ (contributing $\Dow \ppz$). From $\km_{\mbox{\scriptsize MB}} =  (2\freq + \Dow) \ppz -  \omega \jz$ we obtain
 (see Sec. 2 of the Appendix)
\beq
D_{\mbox{\scriptsize eff,MB}} = {\bar{v}^2 \over 2} {\Dow + 2 \freq \over  \omega^2 + \left(\Dow + 2 \freq\right)^2}.
\label{eq:MBmainDeff}
\eeq

\noindent {\bf Run-and-tumble self-propellers (RT)} such as {\it E. coli}~\cite{Berg:2000p601} and Daphnia~\cite{citeulike:11429599,
Ordemann2003260,
Mach2007BMB539} 
undergo intermittent tumbles due to stochastic forces or switching of flagellar beating between the synchronous and asynchronous modes~\cite{Polin-Science:p487, Bennett:2013p586}. They have been studied by {\it ad hoc} models of discrete random walks~\cite{Selmeczi:2005p599,
Selmeczi:2008p603,
citeulike:11429599,
Wu2000EM115}  in quasi-2D with a distribution of turning angles. Although mathematically functional for experimental analysis, such an approach does not unfold the physical processes underlying the motion in continuous time. Here we build a continuous-time model and extend it further to include the effects of engine fluctuations. The velocity direction  $\unitv$ lies in the plane of motion and $\unitw$ represents an axis perpendicular to this plane about which the self-propeller undergoes orientational diffusion (contributing $\Dow \ppz$] and tumbling with frequency $\freq$ [contributing $\freq (1\! -\! \langle \cos\theta\rangle_{\!_P}  ) \ppz\! -\! \freq \langle \sin\theta \rangle_{\!_P} \jz$ to the kinematrix). The kinematrix  $\km_{\mbox{\scriptsize RT}}\! =\! [\Dow\! +\! \freq (1\! - \langle \cos\theta \rangle_{\!_P}  )] \ppz\! -\! \freq \langle \sin\theta \rangle_{\!_P}  \jz$  yields the effective diffusivity for a run-and-tumbler at mean speed $\bar{v}$ (noted by the superscript ``ms''),
\beq
D_{\mbox{\scriptsize eff,RT}}^{\mbox{\scriptsize ms}} = {\bar{v}^2 \over 2} {\Dow +  \freq (1 - \langle\cos\theta\rangle_{\!_P}) \over \left[f\langle\sin\theta\rangle_{\!_P} \right]^2\! +\! \left[\Dow\! +\!  \freq (1 \!-\! \langle\cos\theta\rangle_{\!_P}) \right]^2}
\eeq 
We can extend this model to take into account speed fluctuations $\delta v(t)$ with autocorrelation $\langle \delta v(t)\, \delta v(0)\rangle = [\langle v^2\rangle - \overline{v}^2]e^{-\kappa_v t}$~\cite{Peruani:2007p602}.
Then, the additive modification to effective diffusivity  due to fluctuations (noted by the superscript ``fluc'') is equal to the effective diffusivity of a self-propeller with kinematrix $\km_{\mbox{\scriptsize RT}}^{\mbox{\scriptsize fluc}} = \km_{\mbox{\scriptsize RT}} + \kappa_v{\cal I}$ and mean speed $\sqrt{\langle v^2\rangle - \overline{v}^2}$; that is,
\beq
D_{\mbox{\scriptsize eff,RT}}^{\mbox{\scriptsize fluc}} 
= 
{
\left(\langle v^2\rangle - \bar{v}^2 \right)
\left[\kappa_v + \Dow +  \freq (1 - \langle\cos\theta\rangle_{\!_P})\right]
 \over 
2 \left\{\!
 \left[f\langle\sin\theta\rangle_{\!_P} \right]^2\! +\! \left[\kappa_v\! +\! \Dow\! +\!  \freq (1 \!-\! \langle\cos\theta\rangle_{\!_P}) \right]^2
\! \right\}
 } \,.
\eeq  
Combining these, the effective diffusivity for a run-and-tumbler with speed fluctuations is $D_{\mbox{\scriptsize eff,RT}}^{\mbox{\scriptsize ms}} + D_{\mbox{\scriptsize eff,RT}}^{\mbox{\scriptsize fluc}}$. The same procedure holds for velocity autocorrelation (\ref{eq:angularvelcorrel}) and mean-square-displacement (\ref{eq:MSD1}). We can advance the model further by including more elementary processes in the model, and fitting multiple models in parallel to an experimental dataset to find the best model and elucidate the underlying continuous-time motion of the run-and-tumbler. 

\noindent {\bf 3D self-propeller}: 
The motion of biological and artificial self-propellers in three dimensions is also of interest~\cite{Sandoval2014JFM50, Guizien2006MEPS47,Uttieri2004JPR99}. In contrast to the fixed rotation plane in 2D motion, the plane of rotation for a 3D self-propeller wanders in 3D space. The kinematrix for a self-propeller moving at $\bar{v}$ while rotating at angular speed $\omega$ and suffering orientational diffusion about the three axes of the body frame is $\km_{\mbox{\scriptsize 3D}} = \Dop \ppx + \Dov \ppy + \Dow \ppz - \omega \jz$. Equation (\ref{eq:DeffMinverse}) then yields
\beq
D_{\mbox{\scriptsize eff,3D}} = {\bar{v}^2 \over 3} { \Dow + \Dov  \over \left[ \Dow + \Dop \right]\left[ \Dow + \Dov \right]+\omega^2}.
\eeq 
For a 3D linear motor, setting $\omega =0$
eliminates $\Dov$ from $D_{\mbox{\scriptsize eff,3D}}$, since rotation about $\unitv$ has no observable effect for a linear motor. In that case, $\Dop$ and $\Dow$'s new meanings are  orientational diffusion coefficients about two perpendicular axes orthogonal to $\unitv$.
Although the effects of  rotation-translation coupling in 2D are included implicitly in the phenomenological kinematic parameters, in 3D such hydrodynamic interactions can lead to non-orthogonality of the propulsive velocity and rotation axis. Hence, the velocity has a component $\bar{v}\unitv$ in the instantaneous plane of rotation as well as a component $\bar{v}_\omega\unitw$ along the rotation axis, so the speed is $\sqrt{\bar{v}^2+\bar{v}_\omega^2}$. The effective diffusivity depends on both diagonal and off-diagonal elements of $\km^{-1}$ 
in the form
\begin{align}
D_{\mbox{\scriptsize eff,3D}}^{\mbox{\scriptsize non}\perp} 
=\, 
& {\bar{v}^2 \over d}
\left[\km^{-1}\right]_{22} 
+
{\bar{v} \bar{v}_\omega\over d}
\left[\km^{-1}\right]_{23} 
+
{\bar{v}_\omega \bar{v} \over d}
\left[\km^{-1}\right]_{32} 
\nonumber \\
& +
{\bar{v}_\omega^2 \over d}
\left[\km^{-1}\right]_{33}
.
\end{align} 
For the 3D example here, off-diagonal terms are zero and the correction to
$D_{\mbox{\scriptsize eff,3D}}$ is an additional term 
$
(\bar{v}_\omega^2 /d )
[\km^{-1}]_{33}
=
\bar{v}_\omega^2/3(\Dop + \Dov)
$, which is the effective diffusivity of a 3D linear motor with speed $\bar{v}_\omega$.

\section{UNIVERSALITIES}
The clarity of kinematrix formalism facilitates further insights into universalities that previously had been hiding in the complexities required of differential-equation-based analysis. We can consolidate and recast these four scenarios into a {\it unified} form
\begin{subequations}
\begin{align}
\km_{\mbox{\scriptsize uni}}
& =  \ratew \pz + \ratepv \ppz - \ws \jz +  \delta \left(\px-\py\right)
\\ 
&=
\left[
 \begin{array}{ccc} 
   \ratepv + \delta & \omega_z         &   0  \\ 
   -\omega_z         & \ratepv - \delta  &   0  \\  
   0                        & 0                       &  \ratew
 \end{array} 
 \right]
  \label{eq:matrixformunified}
\end{align}
\label{eq:Mlinearform}
\end{subequations}
where the scenario-dependent parameters are
\beq
\begin{array}{l|cccc}
   & \ratew & \ratepv & \delta & \ws \\
 \hline
\mbox{2D} & 2\freq &\Dow + \freq & \freq & \omega \\
[-2.4 ex] \\
\mbox{RT} & 0 &\! \Dow\! +\! \freq (\!1\! -\! \langle \cos \theta \rangle\!) & 0 & \!\! \freq \langle \sin\theta\rangle  \\
[-2.4 ex] \\
\mbox{MB}& 0 &   \Dow + 2 \freq& 0 & \omega \\
[-2.4 ex] \\
\mbox{3D} & \Dop + \Dov &  \Dow + {\Dop + \Dov \over 2} & {\Dov\! -\! \Dop \over 2} & \omega
\end{array}
\label{eq:parametertable}
\eeq
Using Eqs. (\ref{eq:velcorrel}), (\ref{eq:angularvelcorrel}), (\ref{eq:Mlinearform}) and defining
\beq
\Omega^2 =
 \ws^2 - \delta^2,
\label{eq:capitalOmega}
\eeq
the autocorrelators of angular and linear velocities are
\begin{align}
&
C_{{\bm \omega}{\bm \omega}}(t) = \omega^2  e^{-\ratew t},
\label{eq:omegacord}
\\ 
& 
C_{{\bm v}{\bm v}}(t) = \bar{v}^2  e^{ -\ratepv t} 
  \left\{ \cos\left({\Omega} t \right) + \delta \sin \Omega t /\Omega \right\}
\label{eq:velcord}
\end{align}
{\it Five newly emergent time scales} $\ratew^{-1}$, $\ratepv^{-1}$, $\delta^{-1}$, $\ws^{-1}$, and $\Omega^{-1}$ govern the behavior of all these self-propellers. In 3D $C_{{\bm \omega}{\bm \omega}}$ measures the wandering of the orbital plane, with an exponential decay at characteristic time $\ratew^{-1}$. In 2D it measures the loss of memory of the sense of rotation (i.e. chirality), since $\unitw$ orients to the orbit by a right-hand rule. $C_{{\bm v}{\bm v}}$ measures how fast the velocity forgets its orientation, with characteristic time $\ratepv^{-1}$. The temporal behavior is not governed by $\omega^{-1}$, but $\Omega^{-1}$ which can be real or imaginary depending on $\delta$ and $\ws$. For imaginary $\Omega$,  $C_{{\bm v}{\bm v}}$ does not oscillate but has a longer correlation time $(\ratepv - |\Omega|)^{-1}$ as for linear nanomotors. These time scales also determine the extent to which  $\langle \Delta {\bm r}(\infty) \rangle = (\ratepv^2\! + \Omega^2)^{-1} \bar{v} \!\left[ -\omega \, \unitp(0) +(\delta\! +\!\gamma) \, \unitv(0)\right]$ depends on the direction of initial velocity and, therefore, initial direction of rotation. This is the generalization of the ``chiral diffusion'' of 2D nanorotors~\cite{Nourhani2013p050301}, even though there is no well-defined chirality for a 3D rotary swimmer.

Equations~(\ref{eq:DeffMinverse}) and (\ref{eq:Mlinearform}) yield a unified expression for the effective diffusion coefficient in terms of the new time scales:
\beq
D_{\mbox{\scriptsize eff,uni}} = \left({\bar{v}^2 \over d}\right) \frac{\ratepv + \delta } {\ratepv^2  + \Omega^2}.
\label{eq:genDeff}
\eeq 
We can obtain 
$\Deff$ for any class of self-propeller
 by substituting the appropriate parameters from Eq. (\ref{eq:parametertable}) into Eq.~(\ref{eq:genDeff}). The mean square displacement (\ref{eq:MSD1}) takes a unified form
\begin{align}
\!\!
\langle|\Delta{\bm r}(t)|^2\rangle_{\mbox{\scriptsize uni}}
& = 2 d D_{\mbox{\scriptsize eff}} \, t - 2 \bar{v}^2 \! \left(\ratepv^2\! -\!\Omega^2\! +\! 2 \ratepv \delta\right) \! /\! \left( \ratepv^2\!  +\! \Omega^2 \right)^2
\nonumber \\
& +  {2 \bar{v}^2  e^{-\ratepv t} \over \left( \ratepv^2  + \Omega^2 \right)^2}
\left\{ (\ratepv^2 -\Omega^2 + 2 \ratepv \delta) \cos \Omega t \right.
\nonumber \\
&
\left. \quad  \  \ \ + [(\ratepv^2 - \Omega^2)\delta - 2 \ratepv\Omega^2] \, {\sin \Omega t / \Omega} \right\}\!.
\label{eq:unifiedMSD}
\end{align}
It behaves ballistically (i.e. $\bar{v}^2t^2$) at short times  $t\ll\min (\Omega^{-1} , \ratepv^{-1})$ and diffusively ($2 d D_{\mbox{\scriptsize eff}} \, t$) at long times. For real $\Omega^{-1} \! < \! \ratepv^{-1}$ the crossover between the two limits is oscillatory. However, for $\Omega^{-1} \!>\! \ratepv^{-1}$ the rate at which the velocity forgets its orientation is faster than the oscillation, and the oscillatory crossover is suppressed. For an imaginary $\Omega$, on the other hand, the characteristic time of the exponential decay is $(\ratepv-|\Omega|)^{-1} > \ratepv^{-1}$ with no oscillatory crossover between ballistic and diffusive regimes. 
The special cases of 2D self-propellers with flipping~\cite{Takagi:2013p645} and without flipping~\cite{Ebbens:2010p589} and magnetotactic bacteria \cite{Erglis:2007p596} have been compared thoroughly with numerical experimental data.

\section{CONCLUSION}
The kinematrix formalism elegantly handles a variety of self-propellers with active fluctuations and rotation-translation coupling, and reveals universalities in self-propeller behavior. As a parsimonious means to construct models with different sets of elementary processes, the kinematrix could enable new types of analysis beyond the traditional mode of feeding forward from dynamical model to motor trajectory. For example, distinct stochastic processes that are lumped together in current treatments could be distinguished, such as local environmental noise and internal engine fluctuations which contribute towards a single distribution of turning angles within a traditional discrete model with a fixed-length random walker. In the longer term, memory effects could be disentangled from complex composite behaviors by identifying an optimal memory-free kinematix description and then extracting the residual memory-dependent phenomena. Extension of the formalism to explicitly handle memory effects is also a natural next step; the generality and intuitive clarity of this formalism provides a strong conceptual underpinning to further advances in the analysis of self-propellers.

\section*{ACKNOWLEDGMENT}

This work was supported by the NSF under Grant No. DMR-0820404 through the Penn State Center for Nanoscale Science.

\appendix

\section*{APPENDIX: CALCULATING ENSEMBLE AVERAGE PROPERTIES AND EMERGENT TIME SCALES}
In this appendix we provide a detailed step-by-step walk-through of the application of our theory to two scenarios: the diffusion/flip self-propeller and the magnetotactic bacterium.
We build the kinematrix by inspection and calculate the corresponding effective diffusivity using Eq. (8). We also find the emergent new time scales, $\ratew^{-1}$, $\omega_z^{-1}$, $\ratepv^{-1}$, $\delta^{-1}$, and $\Omega^{-1}$, by comparing the scenaro-specific kinematrix with the general kinematrix form (\ref{eq:Mlinearform}). Then we can calculate the effective diffusivity and mean-square-displacement simply by plugging these parameters into Eqs. (\ref{eq:genDeff}) and (\ref{eq:unifiedMSD}), respectively. \\

\subsection{
Diffusion-Flip self-propellers
\label{sec:App2Dflip}}
The DF self-propeller moves with speed $\bar{v}$ and rotates with angular speed $\omega$ while diffusing orientationally about the axis of rotation $\unitw$ with diffusion coefficient $\Dow$
and flipping at a rate of $\freq$ about the direction of the velocity $\unitv$. The kinematrix includes contributions of three elementary processes:  deterministic rotation about $\unitw$, $\km^{\mbox{\scriptsize det}}_{\unitw} = - \omega {\cal J}_{\omega}$, orientational diffusion about $\unitw$, $\km^{\mbox{\scriptsize ort}}_{\unitw} = D_{\unitw} {\cal P}_\omega^\perp$, and flipping around $\unitv$, $\km^{\mbox{\scriptsize flip}}_{\unitv} = 2 \freq {\cal P}_v^\perp$. As discussed in the body of the paper, we inspect the motion during $[0,dt)$ where ${\cal P}^\perp_{v}=\ppy$,  ${\cal P}^\perp_{\omega}=\ppz$, and ${\cal J}_{\omega} = \jz$. Using the explicit forms of these matrices [Eqs. (\ref{eq:Jmatrices})--(\ref{eq:PPmatrices})], the kinematrix becomes
\begin{align}
\km_{\mbox{\scriptsize 2D}}
& = 
\km^{\mbox{\scriptsize det}}_{\unitw}
+
\km^{\mbox{\scriptsize ort}}_{\unitw}
+
\km^{\mbox{\scriptsize flip}}_{\unitv} 
\nonumber \\
& =   
 -  \omega {\cal J}_\omega
+ \Dow {\cal P}_\omega^\perp
+ 2 \freq  {\cal P}_v^\perp
\nonumber \\
& 
=   
 -  \omega \jz
+ \Dow \ppz 
+ 2 \freq \ppy
\nonumber \\
&=
\left[
 \begin{array}{ccc} 
 \Dow +   2 \freq & \omega        &   0  \\
   -\omega        & \Dow &   0  \\
   0                   & 0                  &   2 \freq
 \end{array} 
 \right]\!\!,
\label{eq:2DKM}
\end{align}
from which we obtain
\beq
\km_{11} \km_{33} - \km_{13}\km_{31} 
=
(2 \freq + \Dow) (2 \freq) - 0 
=
2 \freq (2 \freq + \Dow)
\eeq
and
\beq
 \mbox{det} \km_{\mbox{\scriptsize 2D}}
 =
 2 \freq 
 \left|
 \begin{array}{cc} 
  2 \freq + \Dow & \omega         \\
   -\omega        & \Dow 
 \end{array} 
 \right| 
 =
 2 \freq
 \left[
 \Dow (2 \freq\! +\! \Dow) + \omega^2
 \right] .
\eeq
Employing now Eq. (\ref{eq:DeffMinverse}) yields
\beq
D_{\mbox{\scriptsize eff,2D}}
\!
=
\! 
{\bar{v}^2 \over d} {\km_{11} \km_{33} \!-\! \km_{13}\km_{31} \over \mbox{det} \km}
=
{\bar{v}^2 \over 2}
 {
 2\freq + \Dow 
 \over
 \Dow \left( 2 \freq\! + \!\Dow  \right)
+\omega^2
}
\,
,
\eeq 
which is expression (\ref{eq:Deffective2Dflip}). Comparing the kinematrix (\ref{eq:2DKM}) with the unified kinematrix (\ref{eq:matrixformunified}) immediately yields $\ratew = 2 \freq$, $\omega_z=\omega$, $\ratepv =  \Dow + \freq$, $\delta = \freq$, and $\Omega^2 = \omega^2 - \freq^2$, as tabulated in (\ref{eq:parametertable}). Plugging these parameters into the unified form $\langle|\Delta{\bm r}(t)|^2\rangle_{\mbox{\scriptsize uni}}$ (\ref{eq:unifiedMSD}) gives the mean-square-displacement for diffusion/flip self-propellers,
\begin{widetext}
\begin{align}
\!\!
\langle|\Delta{\bm r}(t)|^2\rangle_{\mbox{\scriptsize 2D}}
&=
2 \bar{v}^2 t
 {
 2\freq + \Dow 
 \over
 \Dow \left( 2 \freq\! +\!\Dow  \right)
\!+\!\omega^2
}
-
2 \bar{v}^2 
 {
 (2\freq + \Dow)^2 - \omega^2 
 \over
\left[ \Dow \left( 2 \freq \!+\!\Dow  \right) \!+\! \omega^2 \right]^2
}
+
2 \bar{v}^2 e^{-(\Dow+\freq)t} 
 {
 (2\freq\! +\! \Dow)^2 - \omega^2 
 \over
\left[ \Dow \left( 2 \freq\! +\!\Dow  \right) \!+\!\omega^2 \right]^2
}
\cos(t\sqrt{ \omega^2\! -\! \freq^2})
\nonumber \\
&
+
2 \bar{v}^2 e^{-(\Dow+\freq)t} \,
 {
 f (2\freq + \Dow)^2 - \omega^2 (3\freq + 2 \Dow)
 \over
\left[ \Dow \left( 2 \freq \!+\!\Dow  \right) +\omega^2 \right]^2
}\,
{\sin(t\sqrt{ \omega^2 - \freq^2}) \over \sqrt{ \omega^2 - \freq^2}}.
\end{align}
\end{widetext}

\subsection{
 Magnetotactic Bacteria 
\label{sec:AppMB}}
In this scenario, motion is restricted to a plane and $\det \km = 0$. 
Magnetotactic bacteria move with speed $\bar{v}$ and rotate in synchrony with a rotating magnetic field at angular speed $\omega$ while suffering orientational diffusion about $\unitw$ and occasional chirality-preserving reversals of $\unitv$ with frequency $\freq$. Three elementary processes contribute to the kinematrix: deterministic rotation about $\unitw$, $\km^{\mbox{\scriptsize det}}_{\unitw} = - \omega {\cal J}_{\omega}$, orientational diffusion about $\unitw$, $\km^{\mbox{\scriptsize ort}}_{\unitw} = D_{\unitw} {\cal P}_\omega^\perp$, and reversals of $\unitv$ with frequency $\freq$ while preserving the chirality which is equivalent to $180^\circ$ rotation of $\hat{v}$ about $\hat{\omega}$, so that $\km^{\mbox{\scriptsize flip}}_{\hat{\omega}} = 2 \freq {\cal P}_\omega^\perp$. Summing up these contributions and keeping in mind that 
${\cal P}^\perp_{\omega}=\ppz$ and ${\cal J}_{\omega} = \jz$,
we obtain the kinematrix for the magnetotactic bacteria, 
\begin{align}
\km_{\mbox{\scriptsize MB}} \!
& = 
\km^{\mbox{\scriptsize det}}_{\unitw}
+
\km^{\mbox{\scriptsize ort}}_{\unitw}
+
\km^{\mbox{\scriptsize flip}}_{\unitw} 
\nonumber \\
& =   
 -  \omega {\cal J}_\omega
+ \Dow {\cal P}_\omega^\perp
+ 2 \freq  {\cal P}_\omega^\perp
\nonumber \\
& 
= \!  
 -  \omega \jz
 \!
+ 
\!
\Dow \ppz 
\!
+
\!
 2 \freq \ppz
\!
\nonumber \\
&=
\!\!
\left[
\!
 \begin{array}{ccc} 
 \Dow +   2 \freq   & \omega               &   0  \\
   -\omega            &\! \Dow + 2 \freq     &   0  \\
   0                       & 0                         &   0
 \end{array} 
 \!
 \right]\!\!.
 \label{eq:MBKM}
\end{align}
In this example, the motion is strictly in the $xy$ plane and $\mbox{det} \, \km_{\mbox{\scriptsize MB}}=0$, such that Eq. (\ref{eq:DeffMinverse}) for effective diffusivity gives the indeterminate form ${0/0}$.
To resolve this problem, as discussed following Eq. (\ref{eq:DeffMinverse}), we replace $\km \to \km^{(\varepsilon)} \equiv \km + \varepsilon {\cal I}$, perform the calculations, and then take the limit $\varepsilon\to0$. Therefore, we write
\beq
\km^{(\varepsilon)}_{\mbox{\scriptsize MB}} = \km_{\mbox{\scriptsize MB}} + \varepsilon {\cal I}
=
\!\!
\left[
\!\!
 \begin{array}{ccc} 
 \Dow\! +\!   2 \freq\! + \!\varepsilon   & \omega               &   0  \\
   -\omega            &\!\! \Dow \!+\! 2 \freq \!+\! \varepsilon    &   0  \\
   0                       & 0                         &    \varepsilon
 \end{array} 
 \!
 \right]\!\!,
\eeq
from which we obtain
\beq
\km_{11}^{(\varepsilon)} \km_{33}^{(\varepsilon)} - \km_{13}^{(\varepsilon)}\km_{31}^{(\varepsilon)} 
=
\varepsilon (\Dow +   2 \freq + \varepsilon)
- 0 
=
\varepsilon (\Dow +   2 \freq + \varepsilon)
\label{eq:MBnomin}
\eeq
and
\begin{align}
 \mbox{det} \km_{\mbox{\scriptsize MB}}^{(\varepsilon)}
 &=
\varepsilon
 \left|
 \begin{array}{cc} 
 \Dow +   2 \freq + \varepsilon & \omega         \\
   -\omega        &  \Dow +   2 \freq + \varepsilon
 \end{array} 
 \right| 
 \nonumber \\
& =
\varepsilon
 \left[
(\Dow +   2 \freq + \varepsilon)^2 + \omega^2
 \right]\!\!.
 \label{eq:MBdenom}
\end{align}
Now, using Eqs. (\ref{eq:DeffMinverse}), (\ref{eq:MBnomin}), and (\ref{eq:MBdenom}) we can calculate the effective diffusivity in the limit $\varepsilon \to 0$,
\begin{align}
D_{\mbox{\scriptsize eff,MB}} 
&= 
\lim_{\varepsilon\to0} {\bar{v} \over 2} { \varepsilon (\Dow +   2 \freq + \varepsilon) \over \varepsilon  \left[ (\Dow +   2 \freq + \varepsilon)^2 + \omega^2 \right] }
\nonumber \\
& =
 {\bar{v}^2 \over 2} {\Dow + 2 \freq \over  \left(\Dow + 2 \freq\right)^2 +  \omega^2}
\end{align}
which is Eq. (\ref{eq:MBmainDeff}). An alternative approach is to compare the kinematrix (\ref{eq:MBKM}) with the unified form (\ref{eq:matrixformunified}) which yields $\omega_z = \omega$, $\ratepv = \Dow + 2 \freq$, $\delta = \ratew = 0$ and $\Omega = \omega$. Plugging these parameters into Eq. (\ref{eq:genDeff}) directly gives Eq. (\ref{eq:MBmainDeff}) without need for the substitution $\km \to \km + \varepsilon {\cal I}$. Moreover, similar to diffusion-flip self-propellers, by plugging these parameters into Eq. (\ref{eq:unifiedMSD}) we obtain the mean-square displacement of magnetotactic bacteria.

\end{document}